# Interactions between Coherent Twin Boundaries and Phase Transition of Iron under Dynamic Loading and Unloading

Kun Wang[a], Jun Chen[a,b*], Xueyang Zhang[c], Wenjun Zhu[d]

[a] *Laboratory of Computational Physics, Institute of Applied Physics and Computational Mathematics, Beijing 100088, PR China*
[b] *Center for Applied Physics and Technology, Peking University, Beijing 100071, China*
[c] *Department of Applied Physics, School of Physics and Electronics, Hunan University, Changsha 410082, China*
[d] *National Key Laboratory of Shock Wave and Detonation Physics, Institute of Fluid Physics, Mianyang 621900, China*

## Abstract

Under high pressures, phase transition, as well as deformation twins, are constantly reported in many BCC metals, whose interactions are of fundamental importance to understand strengthen mechanism of these metals under extreme conditions. However, the interactions between twins and phase transition in BCC metals are remain largely unexplored. In this work, interactions between coherent twin boundaries and α↔ε phase transition of iron are investigated using both non- equilibrium molecular dynamics simulations and nudge elastic band method. Mechanisms of both twin-assisted phase transition and reverse phase transition are studied and orientation relationships between BCC and HCP phase are found to be $\langle 11\bar{1}\rangle_{BCC}||\langle \bar{1}2\bar{1}0\rangle_{HCP}$ and $\langle 1\bar{1}0\rangle_{BCC}||\langle 0001\rangle_{HCP}$ for both cases. The twin boundary corresponds to $\{10\bar{1}0\}_{HCP}$ after the phase transition. It is amazing that the reverse transition seems to be able to "memory" and recover the initial BCC twins. The memory would partly loss when plastic slips take place in the HCP phase before the reverse transition. In the recovered initial BCC twins, three major twin spacing are observed, which are well explained in terms of energy barriers of the transition from HCP phase to BCC twin. Besides, variant selection rule of the twin assisted phase transition is also discussed. The results of present work could be expected to give some clues for producing ultra-fine grain structure in materials exhibiting martensitic phase transition.

## 1. Introduction

Grain boundaries (GBs) have a significant influence on mechanical properties of polycrystalline materials. It is known that dislocation-mediated processes are the main plastic deformation mechanism for coarse-grained materials where GBs could provide obstacles to dislocation slips and, hence, lead to the well-known Hall-Petch relationship between strength and grain size [1, 2]. Whereas, in nanocrystals, inverse Hall-Petch behaviors are widely observed [1], which have been attributed to reasons that the dominate deformation mechanism transfers from dislocation-mediated plasticity to a GB-mediated one. Shear-coupled GB migration, as an effective elementary plastic mechanism at low temperature, has been discussed in details [2, 3]. Besides, experimental data [4] show deformation twinning does also play an important role in ultrafine-grained FCC-Ni whose grain size locates between the two grain sizes mentioned above. Under extreme strain rates, such as shock loadings, the deformation twinning could even occur in coarse-grained materials in FCC metals [5]. This means that the deformation twinning may become a competing deformation mechanism of the dislocation-mediated one, whose interactions with the later one is one of the most important topics in this field. In addition, interactions of dislocations with coherent twin boundaries (CTBs) are an operative deformation mechanism of nanotwinned metals which generally exhibit combinations of excellent mechanical properties, such as high strength and ductility [6-11]. Great progresses have been made in understanding the influence of CTBs on strength and deformation behaviors in nanotwined FCC metals [8, 10, 12]. Because of the existence of twin boundaries, {110}⟨111⟩ slip systems of FCC metals are no longer equivalents. Consequently, different interaction modes emerge, which are referred to as hard mode I (or II) and soft mode according to the degree of inclination between the slip direction (or slip plane) and the twin boundary [11]. Moreover, it is found that the strength does closely relate to twin spacing: The nanotwinned metals would strengthen with the decreasing of twin spacing, but soften after the twin spacing decreasing to a critical value. The strengthening is due to restrictions of dislocation motions provided by CTBs, while the softening results from twin boundary migrations caused by nucleation and propagation of twin dislocations along twin boundary [8, 10]. In contrast, relatively few attentions are paid on the deformation behaviors on nanotwinned BCC metals. A recent study [13] on nanotwinned BCC-iron nanopillars indicates a contrasting role played by CTBs during deformations under tensile and compressive loadings. Under dynamic loadings, deformation twinning is a main mechanism of stress relief at the shock front in many BCC metals, such as Ta [14] and Fe [15, 16]. And phase transitions under high pressures are widely detected or predicted in BCC metals, for example, BCC↔HCP phase transition of iron [17, 18] and Ta [19]. Since both deformation twinning and phase transition are two major local-stress-relief mechanisms in BCC metals, their interactions are of fundamental importance to understand deformation behaviors of these metals under dynamic loadings. However, the interactions between CTB and phase transition of BCC metals remain largely unexplored. This is probably due to two reasons: on one hand, lattice events typically happen at a time scale of ~ ps while picosecond resolution observations at lattice level are forbidden until most recently [20, 21]. On the other hand, studies based on atomic simulations suffer from lack of available interatomic potentials of BCC metals since most semi-/empirical potential of BCC metals developed previously cannot correctly descript phenomena observed at high pressures [14, 22-24]. Although many model potentials (such as bond order potential [25] and generalized pseudopotential theory [26]) have a precision comparable to that of ab-initio calculations, they are not suitable for large-scale and long-time nonequilibrium molecular dynamics (NEMD) simulations because of their computational efficiency. Fortunately, semi-/emperical potentials of some commonly concerned BCC metals are specially developed for high pressure applications, for example, simulating deformation twinning in Ta [14], plasticity and phase transition in Fe [27, 28].

---

[*] E-mail: jun_chen@iapcm.ac.cn (J. Chen)

Studies on interactions between shock induced phase transition and plasticity, including dislocations and twins, have make significant progresses in recent years. Gunkelmann and et. al. [29] studied that the interplay between plasticity and phase transition in nanocrystalline iron and found that the plasticity could happen through dislocation loop emission from GBs before the phase transition. In addition, $(112)[11\bar{1}]$ twins are observed in recovery nanocrystalline iron, which are attributed to shearing along $[11\bar{2}0]_{HCP}$ during reverse transformation from HCP to BCC phase [15, 30]. However, the roles of CTBs played in the phase transition of iron are still not clear. Most recently, researches on CTB and phase transition in Ti has brought out some new insights into deformation mechanisms, as well as its interactions with phase transition, under shock or uniaxial compressions [31-35]. Zong et. al. [33] observed a 90° reorientation of α-Ti before transformation into ω phase under shock compressions by NEMD simulations. The reorientation cannot be attributed to any known twinning modes of HCP metals, but a new deformation mode in extreme conditions, whose formation is accommodated by collective action of dislocations and deformation twins [34]. They further found that it is the reorientation, in contrast to the shear-coupled GB migration mechanism, that contributes to migrations of tilted 90° GB of Ti bicrystals under uniaxial stresses normal to the GB [32]. Zong et. al. [31] also studied the roles of three types of GBs played in the phase transition of Ti under shock compressions, and found that CTB can facilitate nucleation of the phase transition because the CTB is more similar to metastable state in transition path from α to ω phase, and thus helps overcome energy barrier of the phase transition. Despite of these progresses, interactions between CTB and the direction or reverse phase transition at lattice level are still a mystery in BCC metals.

It has been recognized [36] that martensitic phase transition could take place through either stress-assisted and strain induced transition modes (hereafter referred to as SAT and SIT, respectively). The SAT, induced by pressure, nucleates at the preexisting defects, while SIT usually occurs at new lattice defects generated during plastic flow which produces strong local stress concentrations. The SAT of iron is widely observed under static high pressure or shock experiments, but the SIT in iron is not noticed until recent decades [37], which takes place at reduced pressure under pressure and shear operation using rotational diamond anvil cell. Further comparison between the two strain modes has been performed in high-alloy austenitic TRIP steel by Ackermann and et. al [38] using high pressure quasi-hydrostatic experiments recently. Their results indicate that the SIT begins at a lower pressure than that of SAT due to shear stresses deviating from ideal hydrostatic conditions, which supports statements of Levitas [36, 39]. Moreover, the SAT is of block-type and covers complete grains, while the SIT occurs in rather thin deformation bands within austenitic grains. Our recent work [40] on the phase transition of nanocrystalline iron shown that two phase transition modes could occur simultaneously under shock compressions where no macroscopic shear is present. Interestingly, according to our observations [40], BCC grains would completely transform into HCP phase via the SAT, while partially transform via the SIT, which is consistent with experimental observations reviewed in ref. [38]. In the nanocrystalline iron, the SAT is triggered by pressure induced lattice instabilities which will lead to the phase transition taking place in the whole instable grain, while the SIT is induced by high local stress concentrations and will be stamped after the local stresses relieve to a value which is not enough to drive phase interface further. Another difference between the two modes is variant selection of the phase transition. There are quantities of discussions on the variant selection of γ-ε-α' phase transition in various TRIP steels [41, 42], but systematically studies on the variant selections of the phase transition under dynamic loadings are few. Through NEMD simulations on phase transition of both single and nano-/ crystalline iron under shock compressions [28, 40], we found that the variant selection rules of SAT and SIT satisfy strain work criterion and Schmid factor criterion, respectively. Since variants of the SIT nucleate through atom shear or shuffle along specific slip planes (i.e. {110} plane for α↔ε phase transition of iron), inhabited in the phase transition mechanism, during plastic flow represented by dislocation slips, the Schmid factor, as a part of the equation for critical resolved shear stress, is the simplest criteria to judge the active slip systems, and thus determines the selection of the variants. However, the phenomenological variant selection rule for the SAT, i.e., the strain work criterion, is not completely understood. The critical point for the strain work criterion is the definition of stresses used for calculating the strain (or transformation) work. Experimentally, applied stress is usually employed for calculations of the strain (or transformation) work by assuming that local stresses are uniform and equal to the applied stress [41, 42]. In NEMD simulations, the assumption also works well for variant selections of the phase transition of single crystalline iron under shock compressions [28] where local stresses are extremely nonuniform. In similar simulations, but with presence of a cylindrical nanopore in the iron single crystal, the applied stress fails to predict the variant selections of the phase transition nearby the cylindrical nanopore [43]. This may be explained by the local stress concentration nearby the cylindrical nanopore, but when the local stresses, instead of the applied stresses, should be considered is still not clear. Since the latter is more easier to control and measure in experiments, it is of significant importance for material processing and manufacturing to make clear of the conditions when the applied stress is applicable.

In this work, we perform both NEMD simulations to study interactions between CTB and α↔ε phase transition of iron bicrystals under loading and unloading. Mechanisms, as well as the variant selections, of both direct and reverse phase transition are discussed for two cases: the CTB is normal or parallel to the loading direction. And nudge elastic band (NEB) method calculations is employed to interpret the roles of CTBs on the phase transition of iron bicrystals, whose inputs are taken from results of the NEMD simulations. The interpretation could well explain the major twin spacing observed in the NEMD simulations. Results of present studies provide a possible approach to produce ultra-fine grain structure in materials. In the following, simulation methods and sample geometry are described in Section 2. Then results of the NEMD simulations are reported and discussed in Section 3, while further interpretations of the phase transition and reverse phase transition, where BCC twins are present in parent phase and product phase, respectively, are put in Section 4. Finally, we end this work by concluding in Section 5.

## 2. Methodology

Non-equilibrium molecular dynamics (NEMD) simulations are performed by LAMMPS code [44] to investigate α↔ε phase transition of iron bicrystals under ramp compressions. A modified analysis embedded atom method (MAEAM) potential of iron, specially optimized for descriptions of both plasticity and phase transition [28], is employed in this study. Two iron bicrystal samples are constructed: one, with its coherent twin boundary (CTB) normal to Z direction, has a dimension of 18.33×18.20×252.25 nm, and the other, with its CTB normal to X direction (parallel to the Z direction), has an initial size of 28.03×19.82×161.82 nm. Corresponding relationships between crystallographic orientations and coordinate axes of sample reference frame ($\kappa_S$) for each grain in the two bicrystal samples are shown in Fig. 1. For convenience of description, the two samples are referred to as sample I and sample II, respectively. Extra loadings are exerted on the samples through moving an infinite massive piston along Z direction at a speed linearly changing from zero to a certain maximum value $v_p^{max}$ within a ramp rising time $t_{ramp}$. To guarantee sufficient time for the phase transition to proceed nearby the CTBs, the piston sustains at its maximum speed for a time $t_s$ before removing away from compression end of the samples. The simulations are conducted at 0K with a fixed $t_{ramp}$ and $t_s$, both of which is 20 ps in this work, while three maximum particle velocities, i.e., $v_p^{max}$ = 0.4, 0.5, 0.8 km/s and $v_p^{max}$ = 0.5, 0.6, 0.8 km/s, are explored for the two samples, respectively. The whole simulation time is 80ps. In order to check influences of temperature on conclusions arrived at 0K, the two samples with an initial temperature of 300K are employed to repeat the simulations with $v_p^{max}$ = 0.8 km/s. In the NEMD simulations, the simulated systems run in a microcanonical ensemble with a shrink boundary in Z and periodic boundary conditions in the transverse directions. Velocity-Verlet scheme is used to integrate motion equations of atoms, whose timestep is taken to be 0.5 fs. Local lattice structures of simulated samples are identified by adaptive common neighbor analysis method [45] and visualized through Ovito [46] or Atomeye [47]. Wave profiles, represented by local physical quantities (such as particle velocity, local stress and et. al.), are analyzed by 1-D binning of the simulated samples along wave propagation direction. The binning width is 0.785 nm. The local stresses are calculated from atomic simulations based on a continuum-consistent definition [48]. Analysis of c-vector of HCP phase is performed based on method proposed in our previous work [28]. The basic idea of the method is to identify the basal plane of the HCP phase, defined as the plane with maximum atom number, in a cluster centered at each atom with a radius of 3.7Å. Then the c vector at the central atom is taken to be the unit normal to the basal plane, multiplied by magnitude of c-axis which is defined to be double times of distance between two adjacent basal planes. All analyses are performed using a stand-alone post-processing code developed by us.

In atomic simulations, orientation identification of local lattice is essential for studying mechanisms of structural phase transitions of crystals. In present work, the orientation of local lattice in our simulated samples is identified through a lattice tracking method proposed in our previous work [40]. Because the method for identifying the orientation of HCP lattice (mainly c-axis orientation) has mentioned in the last paragraph, we will briefly describe the method of orientation identification for BCC lattice below. The method begins via identifying three crystallographic orientations, i.e., $\frac{1}{2}[111]$, $\frac{1}{2}[\bar{1}11]$ and $\frac{1}{2}[1\bar{1}1]$, whose vectors in $\kappa_S$ are $\boldsymbol{a}_1$, $\boldsymbol{a}_2$ and $\boldsymbol{a}_3$, respectively. For a local lattice element centered at atom $i$, $\boldsymbol{a}_1$, $\boldsymbol{a}_2$ and $\boldsymbol{a}_3$ are three vectors pointing from atom $i$ to its nearest neighbors. It is known that any BCC atom has totally eight vectors, corresponding to eight nearest neighbors. Our scheme to choose the three vectors out of the eight is: 1) Define an initial crystallographic reference frame (ICRF), corresponding to the $\kappa_S$. In this work, our ICRF are ($\boldsymbol{e}_{[111]}^0$, $\boldsymbol{e}_{[1\bar{1}0]}^0$, $\boldsymbol{e}_{[11\bar{2}]}^0$) for sample I and ($\boldsymbol{e}_{[11\bar{2}]}^0$, $\boldsymbol{e}_{[111]}^0$, $\boldsymbol{e}_{[1\bar{1}0]}^0$) for sample II, where $\boldsymbol{e}_{[\cdot]}^0$ denotes unit vector of the corresponding crystallographic orientation. 2) Calculate vectors of $\frac{1}{2}[111]$, $\frac{1}{2}[\bar{1}11]$ and $\frac{1}{2}[1\bar{1}1]$ in the $\kappa_S$, represented by $\mathbf{A}$, through relation of $\mathbf{A} = \mathbf{B}\mathbf{F}^{-1}$, where

$$\mathbf{B} = \begin{bmatrix} 0.5 & 0.5 & 0.5 \\ -0.5 & 0.5 & 0.5 \\ 0.5 & -0.5 & 0.5 \end{bmatrix}, \mathbf{F} = \begin{bmatrix} e^0_{[111]} \\ e^0_{[1\bar{1}0]} \\ e^0_{[11\bar{2}]} \end{bmatrix} \text{ or } \mathbf{F} = \begin{bmatrix} e^0_{[11\bar{2}]} \\ e^0_{[111]} \\ e^0_{[1\bar{1}0]} \end{bmatrix}. \tag{1}$$

3) $\overline{\mathbf{A}} := [\mathbf{a}_1\ \mathbf{a}_2\ \mathbf{a}_3]^T$ is taken to be the one that has the minimum value of $\|\overline{\mathbf{A}} - \mathbf{A}\|$ over all possible permutations, where $\|\cdot\|$ denotes 2-norm of matrix. Then we could obtain any other orientations (for example **b**) of the local lattice in $\kappa_S$ by expressing **b** in terms of $\frac{1}{2}[111]$, $\frac{1}{2}[\bar{1}11]$ and $\frac{1}{2}[1\bar{1}1]$ and further using the obtained coefficient to calculate its vector in $\kappa_S$ (More details can be found in [40]).

A climbing image nudge elastic band method [49] is employed to calculate energy barriers of phase transition of both BCC→HCP and BCC Twin→ HCP. Unit cells of the corresponding phases are taken from results of the NEMD simulations and then expanded periodically to ensure that the atom number of initial and final configuration, used as inputs of the NEB calculations, is equal. More details about the initial and final configurations will be described in Section 4.

## 3. Results

### 3.1 Twin-Activated Phase Transition Mechanism under Ramp Compressions

Under ramp compressions, phase transition of iron is inclined to nucleate at the CTB rather than interior of grains. Detailed scenarios of the phase transition for sample I and II under dynamic loadings are shown in Fig. 2 and Fig. 3, respectively. As shown in the figures, orientation relationships (ORs) between BCC and HCP phase are $[11\bar{1}]_{BCC}||[\bar{1}2\bar{1}0]_{HCP}$ and $[1\bar{1}0]_{BCC}||[0001]_{HCP}$ (or $[111]_{BCC}||[\bar{1}2\bar{1}0]_{HCP}$ and $[10\bar{1}]_{BCC}||[0001]_{HCP}$) when the loading direction is normal to (or parallel to) the CTB. This indicates that the phase transition mechanism, in both cases, is the similar to that happening in iron single crystals [23, 28] where the ORs satisfy $\langle 111 \rangle_{BCC}||\langle \bar{1}2\bar{1}0 \rangle_{HCP}$ and $\langle 10\bar{1} \rangle_{BCC}||\langle 0001 \rangle_{HCP}$. Despite of the similarities in the phase transition mechanism, detailed scenarios of the phase transition for the two cases are apparently different. For the loading parallel to the CTB, the transition products (HCP phase) contains quantities of slip bands (represented by green atoms in Fig. 2) which result from activations of $\{011\}\langle 111 \rangle$ slips in the BCC phase (or $\{0001\}\langle \bar{1}2\bar{1}0 \rangle$ slips in the HCP phase). In contrast, a defect-free HCP phase is obtained through the phase transition at the CTB normal to the loading direction. However, in this case, the slip bands may emerge in HCP phase generated through phase transition occurring at the interior of Grain I when $v_p^{max} > 0.5$ km/s (See Supplementary Materials). As a consequence, heterogeneous defects are introduced into the defect-free HCP phase after the Grain I is completely transformed into HCP phase, which will result in formations of secondary twins during the unloading (See Fig. 4). At 300 K, similar processes are observed but with less time (See Supplementary Materials). Another difference is variant selections of the phase transition. Different variants could be distinguished by c-vector of the HCP phase. Through method mentioned in *Methodology*, the c-vectors are calculated for all HCP atoms and then used to construct pole figures for the two cases. From results shown in Fig. 5, we find that only one variant, i.e., $(1\bar{1}0)_{BCC}$, is generated, while $(10\bar{1})_{BCC}$ and $(011)_{BCC}$ variant emerges in Grain I and Grain II, respectively. The results of variant selections will be further summarized in terms of a phenomenological rule in the next part. Besides, the c-vector, i.e., $[0001]_{HCP}$ direction, is determined to point along $[1\bar{1}0]_{BCC}$ and $[10\bar{1}]_{BCC}$ in Grain I for the two samples, respectively, which confirms the results analyzed in Fig. 2 and Fig. 3.

Studies of the dynamics of the phase transition under dynamic loadings need to know not only the phase transition mechanism, but also onset of the phase transition or transition criteria, as well as energy barrier of the transition. The energy barrier will be discussed in Section 4. Below, the onset of the phase transition, activated by the CTB normal to the loading direction, is analyzed by monitoring evolutions of both microstructures and wave profiles. According to our results (See Fig. 6 and Supplementary Materials), the phase transition begins, for example, at 30ps for the loading with $v_p^{max} = 0.4$ km/s and at 28ps for the loading with $v_p^{max} = 0.4$ km/s. Before the onset of the phase transition, a "jump" in elastic precursor could be observed during wave propagations (See Fig. 6) and the phase transition happens at exact time when the "jump" arrives at the position of the CTB. And this phenomenon is also observed in all situations explored in this work, including different applied strain rates and initial temperature. From our recent studies [50] on the phase transition of iron single crystals under ramp compressions, we known that the "jump" in the elastic precursor is due to mechanical instabilities induced by strain gradients during compressions. Apparently, a sustained strain gradient in Grain I is created by the ramp compressions, and this strain gradient will be influenced when the elastic precursor approaches to the CTB. Thus, it could be concluded that the phase transition begins when the strain-gradient induced instabilities, due to additional strain gradient disturbances applied by the CTB on the sustained strain gradient "state", take place.

## 3.2 Variant Selection Rule of the Twin-Activated Phase Transition

Variant selections, as a result of the phase transition, determine microstructures and textures in the transitioned materials, and thus influence mechanical behaviors of the materials. As reviewed in the *Introduction*, two criteria, i.e., the Schmid factor and strain (or transformation) work, are found to be generally applied for variant selections of a serial of phase transitions. Under dynamic loading, they seem to associate with the strain-induced and stress-assisted phase transition mode (SIT and SAT), respectively, at least for iron [40, 43]. We will check which criterion is obeyed by the CTB activated phase transition.

In accordance with the scheme mentioned in ref. [40], deformation gradient due to the phase transition of iron, in crystallographic reference frame ($\kappa_\alpha$), i.e., $(\mathbf{e}_{[1\bar{1}0]}, \mathbf{e}_{[110]}, \mathbf{e}_{[001]})$, is

$$\boldsymbol{F}_\alpha = \begin{bmatrix} 1 & 0 & 0 \\ 0 & 3\sqrt{2}/4 & 0 \\ 0 & 0 & \sqrt{3}/2 \end{bmatrix}. \tag{2}$$

It could be transformed into the $\kappa_S$ by

$$\boldsymbol{F}_S = \boldsymbol{\mathcal{T}}_{\alpha \to S}^{-1} \boldsymbol{F}_\alpha \boldsymbol{\mathcal{T}}_{\alpha \to S} \tag{3}$$

where $\boldsymbol{\mathcal{T}}_{\alpha \to S}$ is transition matrix from the $\kappa_\alpha$ to the $\kappa_S$. According to the strain work criterion, transformation work per volume is

$$W = \frac{1}{2}\boldsymbol{\sigma} : \boldsymbol{\varepsilon}_S \tag{4}$$

where $\boldsymbol{\sigma}$ is the local stress nearby the CTB at the time just before the phase transition, and transition strain relates to the deformation gradient by

$$\boldsymbol{\varepsilon}_S = \frac{1}{2}(\boldsymbol{F}_S + \boldsymbol{F}_S^T) - \boldsymbol{I}, \tag{5}$$

where $\boldsymbol{I}$ is the unit tensor. Thereby, to evaluate the transformation work, we need known $\boldsymbol{\sigma}$ and the vectors of $\{\mathbf{e}_{[1\bar{1}0]}, \mathbf{e}_{[110]}, \mathbf{e}_{[001]}\}$ in $\boldsymbol{\kappa}_S$. Based on the method mentioned in Section 2, the local stresses are calculated to be

$$\boldsymbol{\sigma}^N = \begin{bmatrix} 11.86 & 0.00 & 0.00 \\ 0.00 & 6.63 & 0.00 \\ 0.00 & 0.00 & 18.81 \end{bmatrix} \text{ and } \boldsymbol{\sigma}^P = \begin{bmatrix} 10.64 & 0.00 & 0.00 \\ 0.00 & 5.83 & 0.00 \\ 0.00 & 0.00 & 16.71 \end{bmatrix}, \tag{6}$$

for sample I under the loading with $v_p^{max} = 0.4$ km/s, and for sample II under the loading with $v_p^{max} = 0.6$ km/s, respectively, where units are GPa. While the vectors of the crystallographic orientations are analyzed using method mentioned in our previous publication [40]. The calculated transformation works for all possible variants are listed in Table I. Besides, the Schmid factors with respective to all {110}⟨111⟩ slip systems of BCC iron are also calculated with the determined local stresses and the vectors of crystallographic orientations in $\boldsymbol{\kappa}_S$ (See Table II). According to the results shown in the two tables, maximum transformation work criterion is obeyed by the twin activated phase transitions in both samples. That is to say, the twin activated phase transition of iron priors to happen through the SAT mode. This may be because that the lack of interface dislocations in the CTB leads to relatively uniform local stresses distributed in cross sections of the simulated samples, and thus facilitate the phase transition through the SAT mode just like that happens in single crystals. Interestingly, for the Sample I, variant with the maximum transformation work only appears in Grain II although the variant in Grain I is the same as the one in Grain II. This means that the phase transition first takes place in Grain II and then proceeds in both grains via phase interfaces. In contrast, both two observed variants have maximum transformation work in each grain of sample II. This means that each grain of sample II will have two types of variants after the phase transition, which is consistent with results from NEMD simulations. Additionally, we further calculate the transformation work using a reduced applied stress —— uniaxial stress, defined by

$$\sigma_a = \begin{bmatrix} 0 & 0 & 0 \\ 0 & 0 & 0 \\ 0 & 0 & 1 \end{bmatrix}. \tag{7}$$

From the calculation results listed in Table III, the same conclusions could be arrived as that from Table I within ranges of allowable error. This result suggests that the "shape" (uniaxiality) of the applied stress is nearly not affected by the CTBs before the twin- activated phase transition under dynamic loadings.

3.3 Twins Generated by Reverse Transition from HCP to BCC upon Unloading

Reverse transition from HCP to BCC phase are observed upon unloading from the compressed end of the simulated samples and quantities of twins are residue in recovery samples after the unloading (See Fig. 7). The unloading wave in the HCP phase propagates faster than the speed of phase interface between HCP and BCC phase and reverse transitions could happen almost simultaneously in multiple positions along the wave propagation direction (See Fig. 4 marked by blue arrows). Mechanism of the reverse transition of iron at atom level has been shown in Fig. 8, which is attributed to the shear along $\{10\bar{1}0\}_{HCP} \parallel \{11\bar{2}\}_{BCC}$ planes. This mechanism is also reported by Wang and et. al. through inferred from microstructures retained in the post-shock iron sample [15]. According to the mechanism, the reverse transition could generate three sets of {112} twins with a three-fold symmetry. However, the reverse transition seems to be able to "memory" the initial twin in BCC samples, and tends to generate the twin first (See Fig. 7). The other two kinds of twins, for example, the secondary twins (marked by yellow arrows in Fig. 4d and 4e) in sample I and hexagon-shaped twins (marked by yellow arrows in Fig. 7b) in sample II, would be generated only if plastic slips happen in the HCP phase before the reverse transition. In this term, the plastic slips are the reason for the broken of the memory.

From the discussion above, only one type of twins would be generated by the reverse transition when no plastic slips occur in the HCP phase. This is indeed the case as observed in sample I, recovered for the loading with $v_p^{max}$ = 0.4 km/s or $v_p^{max}$ = 0.5 km/s, where no plastic slips take place in the HCP phases after the phase transition (See supplementary materials). From Fig. 7, we find that the original BCC twin is refined significantly after a transition circle of α→ε→α. This may provide new clues for present boundary engineering to manufactory novel materials with ultrafine twins. Distribution of twin spacing in recovery sample from the loading with $v_p^{max}$ = 0.4 km/s is drawn in Fig. 9, which shows three major twin spacing, i.e., $d_0$, $2d_0$ and $3d_0$ ($d_0$ is the separation between two adjacent {112} planes). The reasons for the distribution in twin spacing could be explained by energy barriers between HCP phase and the twin with different twin spacing, which will discussed further in the next section.

## 4. Discussions

Energy barriers of the phase transitions, including BCC ↔ HCP and BCC twin ↔ HCP, are calculated using NEB method. According to the ORs between BCC and HCP phase shown in Fig. 2, initial and the corresponding final configurations for the twin-activated phase transition of iron could be obtained, which have dimensions of $d_{[11\bar{1}]} \times 2d_{[112]} \times d_{[1\bar{1}0]}$ and $d_{[\bar{1}2\bar{1}0]} \times 3d_{[\bar{1}010]} \times d_{[0001]}$ along X, Y and Z, respectively, where $d_{[\cdot]}$ denotes the minimum lattice periodicity along the corresponding crystallographic direction. For brevity, we refer to the transition of BCC ↔ HCP as an elementary transition. To calculate the energy barrier between BCC twin and HCP phase, the initial BCC configuration should be modified to contain CTBs without changing the ORs. This is realized through dividing the initial configuration into two equivalent unit cells, sizes of $d_{[11\bar{1}]} \times d_{[112]} \times d_{[1\bar{1}0]}$, and then substituting the second cell into one with a dimension of $d_{[111]} \times d_{[\bar{1}\bar{1}2]} \times d_{[1\bar{1}0]}$. Thus, we have obtained an elementary transition of BCC-twin ↔ HCP. Through repeating the two cell along –Y and +Y direction by $n_{tw}$ times, respectively, two BCC-twins with twin spacing of $n_{tw}d_{\langle 112\rangle}$ could be obtained. And, accordingly, the final configuration for HCP phase, repeated along Y direction by $n_{tw}$ times, serves as new final configuration. The resulting configurations at initial and final states for the case of $n_{tw} = 2$ are shown in Fig. 10. Linear interpolation between the initial and final configurations is employed to construct intermediate images as our initial input, where corresponding relationships between atoms in the initial and final configurations should be given. In present work, the corresponding relationships are established via minimal global displacement principle, generally employed to infer the Bain mechanisms for martensitic phase transitions. Totally 40 images along minimum energy path (MEP) are involved and each image relaxes for 500 ps under periodic boundary conditions along X, Y and Z.

As shown in Fig. 11, energy barriers for α↔ε phase transition of iron are defined as the maximum height between adjacent minima and maxima along the direct or reverse MEP. Similarly, energy barrier for the transition between HCP phase and BCC-twins with different twin spacing could be evaluated. From results shown in Fig. 12, energy barrier of the transition from BCC-twin with a twin spacing of $2d_{\langle 112 \rangle}$ or $3d_{\langle 112 \rangle}$ to HCP phase is lower than that from BCC to HCP phase. This means that the phase transition of iron with BCC-twins is more energetically favorable than that of single crystals, which supports the observations in Section 3 where α→ε phase transition prior to nucleation at the CTB.

Below, we will consider the most energetically favorable twin spacing of the reverse transition induced twins observed in our simulations. However, it is not appropriate to compare among energy barriers of the transition from HCP phase to BCC-twins of different twin spacing directly, since all the cases of different twin spacing contain two elemental transitions of HCP → BCC twin and they are only differed in the number of the elemental transition of HCP → BCC. Thus, to determine which BCC-twin is more energetically favorable during the reverse transition, the influence of the additional elemental transition of HCP → BCC should be excluded before the comparisons. A way to exclude the additional elemental transition is to define another energy function, different from the potential energy ($E^d_{\varepsilon \to TW}$) along MEP of transition from HCP to BCC-twin, namely,

$$E^d(R) = n_{tw}E^d_{\varepsilon \to TW}(R) - (n_{tw} - 1)E_{\varepsilon \to \alpha}(R), \tag{8}$$

where $R$ represents reaction coordinate, $E_{\varepsilon \to \alpha}$ is the potential energy along MEP of the elementary transition of HCP → BCC, and the upper index $d$ is used to distinguish BCC-twins of different twin spacing. With the definition (8), energy barriers on $E^d(R)$ are calculated for cases of different twin spacing (See Fig. 13). From the calculated results, BCC-twins with a twin spacing of $2d_{\langle 112 \rangle}$ is the most energetically favorable and the ones with a twin spacing of $d_{\langle 112 \rangle}$ of $3d_{\langle 112 \rangle}$ have an energy very close to the most energetically favorable one. And the energy barrier, to form BCC-twins with a twin spacing larger than $3d_{\langle 112 \rangle}$, is larger than the most energetically favorable one by at least two orders of magnitude. Thus, we could infer that BCC twins with a twin spacing of $d_{\langle 112 \rangle}$, $2d_{\langle 112 \rangle}$ and $3d_{\langle 112 \rangle}$ are the most likely products of transition from HCP to BCC twin, which agrees well with the statistic results from NEMD simulations.

## 5. Conclusions

In conclusion, we have performed both NEMD simulations and nudge elastic band (NEB) method calculations to study the roles of CTBs played on α↔ε phase transition of iron bicrystals under loading and unloading. Two typical bicrystal samples, with its CTB normal or parallel to the loading directions, are employed. Results of the NEMD simulations show that phase transition, activated by CTB, takes place at a pressure lower than that in single crystals. Different from the phase transition triggered by dislocations [37] or high angle grain boundaries in nanocrystals [40], the CTB activated phase transition takes place via the SAT mode whose variant selection rule obeys maximum transformation work criterion. And applied stress is found to be as effective as local stress in determination of the variant selections for the CTB activated phase transition, which is unexpected since lattice defects would create local stress concentrations and lead to screening effects (unaxiality) of the applied stress nearby the defects. This result indicates that the local stress concentration generated by the CTB is too small to affect the effects of the applied stress under dynamic loadings. Mechanisms for both direct and reverse phase transition are studied in terms of orientation relationships between BCC and HCP phase, which are $\langle 11\bar{1}\rangle_{BCC}||\langle\bar{1}2\bar{1}0\rangle_{HCP}$ and $\langle 1\bar{1}0\rangle_{BCC}||\langle 0001\rangle_{HCP}$. The CTB of the BCC phase, i.e., $\{112\}_{BCC}$, corresponds to $\{10\bar{1}0\}_{HCP}$ after the phase transition. Besides, quantities of twins are generated during reverse transition. According to the mechanism of the reverse phase transition, three sets of $\{112\}$ twins with a three-fold symmetry would be generated equivalently. However, the initial twin is the major product after the reverse transition, while the other two sets of twins could be observed only if plastic slips take place in the HCP phase before the reverse transition. And twin spacing of the major twins mainly distributes at three values, i.e., $d_{\langle 112\rangle}$, $2d_{\langle 112\rangle}$ and $3d_{\langle 112\rangle}$, where $d_{\langle 112\rangle}$ is one lattice periodicity along $\langle 112\rangle$ direction. Through NEB calculations, we find that twins with the three spacing are more energetically favorable in energy barriers than that with larger twin spacing. The results of present study provide a possible approach to produce ultra-fine grain structure in materials.

# Acknowledgements

This work is supported by the National Natural Science Foundation of China (NSFC-NSAF 11076012) and China Postdoctoral Science Foundation (No. 2017M610824).

# References

[1] R.J. Asaro, S. Suresh, Acta Materialia, 53 (2005) 3369-3382.
[2] A. Karma, Z.T. Trautt, Y. Mishin, Physical Review Letters, 109 (2012) 095501.
[3] A. Rajabzadeh, F. Mompiou, M. Legros, N. Combe, Physical Review Letters, 110 (2013) 265507.
[4] S. Cheng, A.D. Stoica, X.L. Wang, Y. Ren, J. Almer, J.A. Horton, C.T. Liu, B. Clausen, D.W. Brown, P.K. Liaw, L. Zuo, Physical Review Letters, 103 (2009) 035502.
[5] F. Zhao, L. Wang, D. Fan, B.X. Bie, X.M. Zhou, T. Suo, Y.L. Li, M.W. Chen, C.L. Liu, M.L. Qi, M.H. Zhu, S.N. Luo, Physical Review Letters, 116 (2016) 075501.
[6] L. Zhang, C. Lu, K. Tieu, Computational Materials Science, 118 (2016) 180-191.
[7] F. Sansoz, K. Lu, T. Zhu, A. Misra, MRS Bulletin, 41 (2016) 292-297.
[8] D. Jang, X. Li, H. Gao, J.R. Greer, Nat Nano, 7 (2012) 594-601.
[9] J. Li, J.Y. Zhang, G. Liu, J. Sun, International Journal of Plasticity, 85 (2016) 172-189.
[10] X. Zhao, C. Lu, A.K. Tieu, L. Pei, L. Zhang, L. Su, L. Zhan, Materials Science and Engineering: A, 687 (2017) 343-351.

[11] T. Zhu, H. Gao, Scripta Materialia, 66 (2012) 843-848.
[12] I.A. Ovid'Ko, A.G. Sheinerman, Reviews on Advanced Materials Science, (2016).
[13] G. Sainath, B.K. Choudhary, Philosophical Magazine, 96 (2016) 3502-3523.
[14] R. Ravelo, T.C. Germann, O. Guerrero, Q. An, B.L. Holian, Physical Review B, 88 (2013) 134101(134117 pp.)-134101(134117 pp.).
[15] S.J. Wang, M.L. Sui, Y.T. Chen, Q.H. Lu, E. Ma, X.Y. Pei, Q.Z. Li, H.B. Hu, Sci. Rep., 3 (2013).
[16] L.M. Dougherty, G.T. Gray Iii, E.K. Cerreta, R.J. McCabe, R.D. Field, J.F. Bingert, Scripta Materialia, 60 (2009) 772-775.
[17] T. Takahashi, W.A. Bassett, Science, 145 (1964) 483-486.
[18] D. Bancroft, E.L. Peterson, S. Minshall, Journal of Applied Physics, 27 (1956) 291-298.
[19] C.H. Lu, E.N. Hahn, B.A. Remington, B.R. Maddox, E.M. Bringa, M.A. Meyers, Scientific Reports, 5 (2015) 15064.
[20] N.J. Hartley, N. Ozaki, T. Matsuoka, B. Albertazzi, A. Faenov, Y. Fujimoto, H. Habara, M. Harmand, Y. Inubushi, T. Katayama, M. Koenig, A. Krygier, P. Mabey, Y. Matsumura, S. Matsuyama, E.E. McBride, K. Miyanishi, G. Morard, T. Okuchi, T. Pikuz, O. Sakata, Y. Sano, T. Sato, T. Sekine, Y. Seto, K. Takahashi, K.A. Tanaka, Y. Tange, T. Togashi, Y. Umeda, T. Vinci, M. Yabashi, T. Yabuuchi, K. Yamauchi, R. Kodama, Applied Physics Letters, 110 (2017) 071905.
[21] R. Briggs, M.G. Gorman, A.L. Coleman, R.S. McWilliams, E.E. McBride, D. McGonegle, J.S. Wark, L. Peacock, S. Rothman, S.G. Macleod, C.A. Bolme, A.E. Gleason, G.W. Collins, J.H. Eggert, D.E. Fratanduono, R.F. Smith, E. Galtier, E. Granados, H.J. Lee, B. Nagler, I. Nam, Z. Xing, M.I. McMahon, Physical Review Letters, 118 (2017) 025501.
[22] K. Kadau, T.C. Germann, P.S. Lomdahl, R.C. Albers, J.S. Wark, A. Higginbotham, B.L. Holian, Physical Review Letters, 98 (2007) 135701.
[23] K. Kadau, T. Germann, P. Lomdahl, B. Holian, Physical Review B, 72 (2005) 064120.
[24] A. Higginbotham, M.J. Suggit, E.M. Bringa, P. Erhart, J.A. Hawreliak, G. Mogni, N. Park, B.A. Remington, J.S. Wark, Physical Review B, 88 (2013) 104105.
[25] D.G. Pettifor, Physical Review Letters, 63 (1989) 2480-2483.
[26] J.A. Moriarty, L.X. Benedict, J.N. Glosli, R.Q. Hood, D.A. Orlikowski, M.V. Patel, P. Söderlind, F.H. Streitz, M. Tang, L.H. Yang, Journal of Materials Research, 21 (2006) 563-573.
[27] N. Gunkelmann, E.M. Bringa, D.R. Tramontina, C.J. Ruestes, M.J. Suggit, A. Higginbotham, J.S. Wark, H.M. Urbassek, Physical Review B, 89 (2014) 140102.
[28] K. Wang, S. Xiao, H. Deng, W. Zhu, W. Hu, International Journal of Plasticity, 59 (2014) 180-198.
[29] N. Gunkelmann, E.M. Bringa, H.M. Urbassek, Journal of Applied Physics, 118 (2015) 185902.
[30] N. Gunkelmann, D.R. Tramontina, E.M. Bringa, H.M. Urbassek, Journal of Applied Physics, 117 (2015) 085901.
[31] H. Zong, X. Ding, T. Lookman, J. Sun, Acta Materialia, 115 (2016) 1-9.
[32] H. Zong, X. Ding, T. Lookman, J. Li, J. Sun, Acta Materialia, 82 (2015) 295-303.
[33] H. Zong, T. Lookman, X. Ding, S.-N. Luo, J. Sun, Acta Materialia, 65 (2014) 10-18.
[34] H. Zong, X. Ding, T. Lookman, J. Li, J. Sun, E.K. Cerreta, J.P. Escobedo, F.L. Addessio, C.A. Bronkhorst, Physical Review B, 89 (2014) 220101.
[35] S. Rawat, N. Mitra, Computational Materials Science, 126 (2017) 228-237.
[36] V.I. Levitas, Physical Review B, 70 (2004) 184118.
[37] Y. Ma, E. Selvi, V.I. Levitas, J. Hashemi, Journal of Physics: Condensed Matter, 18 (2006) S1075.
[38] S. Ackermann, S. Martin, M.R. Schwarz, C. Schimpf, D. Kulawinski, C. Lathe, S. Henkel, D. Rafaja, H.

Biermann, A. Weidner, Metallurgical and Materials Transactions A, 47 (2016) 95-111.

[39] V.I. Levitas, EPL (Europhysics Letters), 66 (2004) 687.

[40] K. Wang, W. Zhu, S. Xiao, K. Chen, H. Deng, W. Hu, International Journal of Plasticity, 71 (2015) 218-236.

[41] N. Li, Y.D. Wang, W.J. Liu, Z.N. An, J.P. Liu, R. Su, J. Li, P.K. Liaw, Acta Materialia, 64 (2014) 12-23.

[42] M. Humbert, B. Petit, B. Bolle, N. Gey, Materials Science and Engineering: A, 454–455 (2007) 508-517.

[43] L. Wu, K. Wang, S. Xiao, H. Deng, W. Zhu, W. Hu, Computational Materials Science, 122 (2016) 1-10.

[44] S. Plimpton, Journal of Computational Physics, 117 (1995) 1-19.

[45] A. Stukowski, Modelling and Simulation in Materials Science and Engineering, 20 (2012) 045021.

[46] A. Stukowski, Modelling and Simulation in Materials Science and Engineering, 18 (2010).

[47] J. Li, Modelling and Simulation in Materials Science and Engineering, 11 (2003) 173-177.

[48] J. Cormier, J.M. Rickman, T.J. Delph, Journal of Applied Physics, 89 (2001) 99-104.

[49] G. Henkelman, B.P. Uberuaga, H. Jónsson, The Journal of Chemical Physics, 113 (2000) 9901-9904.

[50] K. Wang, J. Chen, W. Zhu, W. Hu, M. Xiang, International Journal of Plasticity, (2017).

Table I. Calculated transformation works for all possible martensitic variants represented by its basal plane. Sample I (or II) is under the loading along a direction normal (or parallel) to the CTB with $v_p^{max}$ = 0.4 (or 0.6) km/s. Actual variants observed in our simulations are colored by red.

| Variant | Sample I | | Sample II | |
|---|---|---|---|---|
| | Grain I | Grain II | Grain I | Grain II |
| (110) | 9.86126 | 3.55717 | 11.0212 | 10.1666 |
| (101) | 10.3001 | 4.04662 | 9.00754 | 10.0200 |
| (011) | 8.3276 | 7.59849 | 11.1011 | *10.9613* |
| (1$\bar{1}$0) | *9.05835* | *26.8654* | 4.64198 | 7.13025 |
| (10$\bar{1}$) | 6.65681 | 11.1415 | *11.2677* | 9.76641 |
| (01$\bar{1}$) | 7.27849 | 3.29344 | 9.84491 | 7.84257 |

Table II. Calculated Schmid factors with respective to all possible {110}⟨111⟩ slip systems of BCC iron. Other conventions are the same as Table I.

| Variant | Slip direction | Sample I | | Sample II | |
|---|---|---|---|---|---|
| | | Grain I | Grain II | Grain I | Grain II |
| (110) | [1$\bar{1}$1] | 0.09909 | 0.61435 | 0.15687 | 0.54080 |
| | [1$\bar{1}\bar{1}$] | 0.09909 | 0.23638 | 0.16625 | 0.54221 |
| (101) | [11$\bar{1}$] | 0.13217 | 0.55106 | 0.12538 | 0.06055 |
| | [1$\bar{1}\bar{1}$] | 0.24513 | 0.51831 | 0.29270 | 0.60054 |
| (011) | [1$\bar{1}$1] | 0.24513 | 0.65332 | 0.29029 | *0.60045* |
| | [11$\bar{1}$] | 0.13217 | 0.24450 | 0.12950 | *0.06031* |
| ($\bar{1}$10) | [111] | *0.00000* | *0.71963* | 0.00193 | 0.00076 |
| | [11$\bar{1}$] | *0.00000* | *0.31451* | 0.00270 | 0.00020 |
| (10$\bar{1}$) | [111] | 0.00511 | 0.53253 | *0.00896* | 0.47376 |
| | [1$\bar{1}$1] | 0.25943 | 0.40360 | *0.15593* | 0.03735 |
| (01$\bar{1}$) | [111] | 0.00511 | 0.75332 | 0.01147 | 0.47266 |
| | [1$\bar{1}\bar{1}$] | 0.25943 | 0.19110 | 0.15298 | 0.03773 |

Table III. The same as Table I, except for using uniaxial stress, instead of local stress, to calculate the transformation works.

| Variant | Sample I | | Sample II | |
| --- | --- | --- | --- | --- |
| | Grain I | Grain II | Grain I | Grain II |
| (110) | 0.41570 | 0.89846 | 0.35530 | 0.35608 |
| (101) | 0.45263 | 0.23523 | 0.32030 | 0.32136 |
| (011) | 0.29017 | 0.87518 | 0.48061 | *0.48198* |
| (1$\bar{1}$0) | *0.16612* | *1.31598* | 0.17772 | 0.17813 |
| (10$\bar{1}$) | 0.04972 | 0.03482 | *0.48331* | 0.48262 |
| (01$\bar{1}$) | 0.01778 | 0.79239 | 0.32096 | 0.32165 |

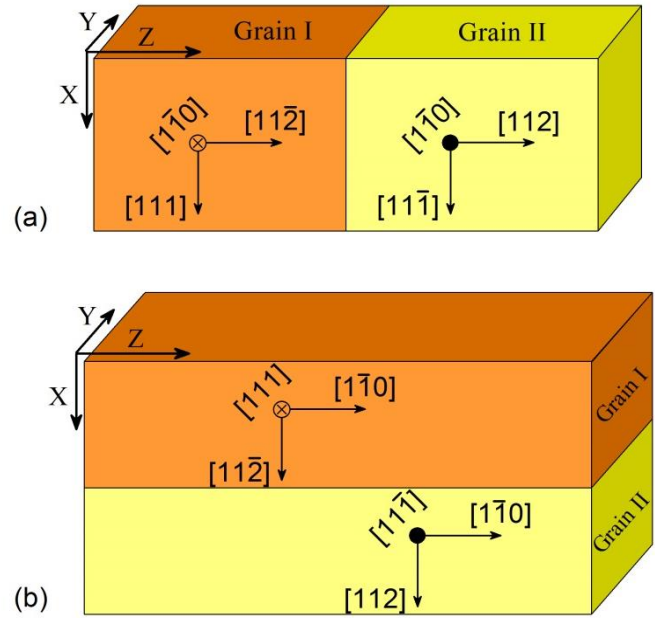

Fig. 1. Schematics of iron bicrystals (a) sample I and (b) II employed in present study, where waves propagate along +Z direction. Relationships between sample reference frames and crystallographic reference frames are shown in the figure for both Grain I and II.

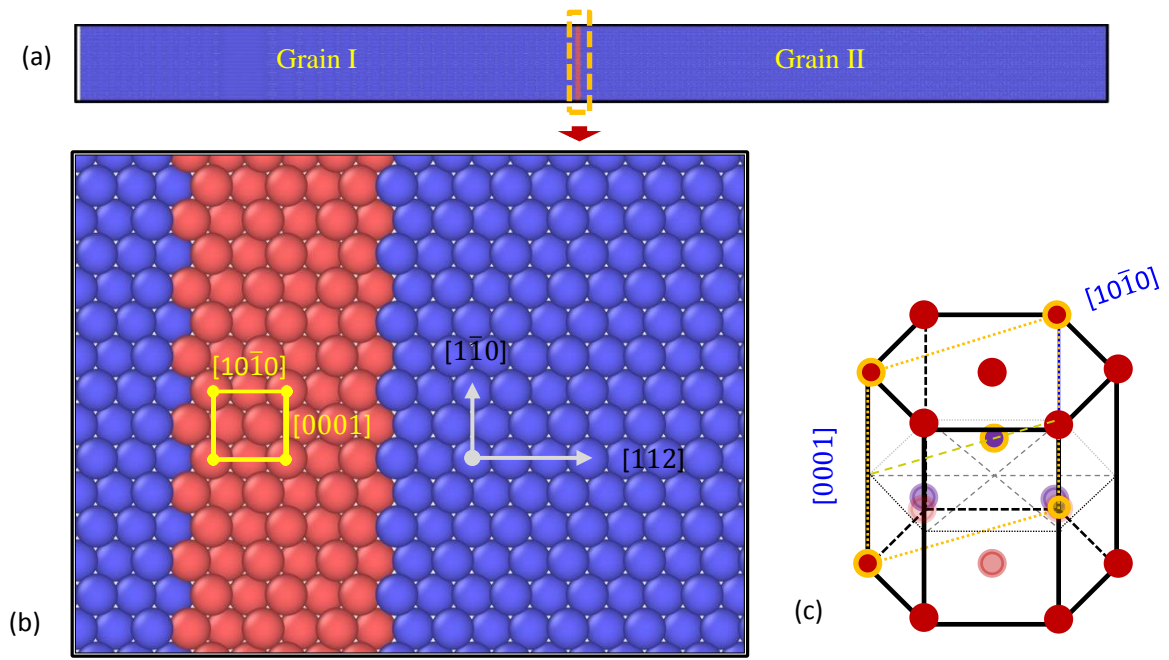

Fig. 2. (a) Snapshot of sample I at 31ps under the ramp with $v_p^{max}$ = 0.4 km/s, where the region marked by dash square has been amplified in (b), and (c) a typical cell of HCP iron. Common neighbor analyses have been employed to identify local structures in the snapshot: blue (BCC), yellow (FCC), red (HCP) and gray (lattice defects). A unit cell of HCP iron is delineated by yellow solid lines in the figure (b), where the delineated atom plane is corresponding to the plane consisting of yellow atoms shown in the figure (c). Orientation relationships between BCC and HCP iron are identified to be $[11\bar{1}]_{BCC}||[\bar{1}2\bar{1}0]_{HCP}$ and $[1\bar{1}0]_{BCC}||[0001]_{HCP}$. Besides, the CTB of BCC phase, i.e., $(112)_{BCC}$ corresponds to $(10\bar{1}0)_{HCP}$ after the phase transition.

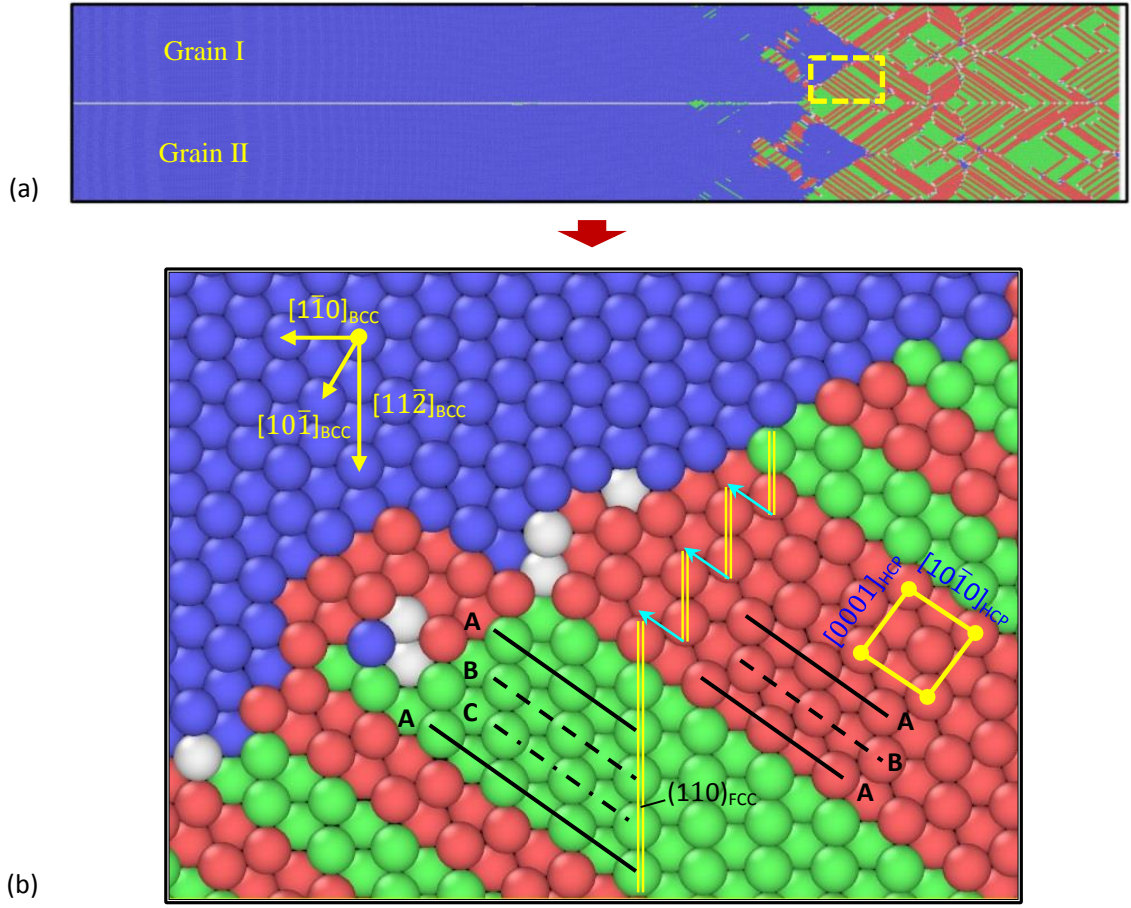

Fig. 3. (a) Snapshot of sample II at 30ps under ramp compressions with $v_p^{max}$ = 0.6 km/s, where the region circled by the yellow dashed square are amplified in (b). Colors assigned to each atom have the same meaning as Fig. 2. In figure (a), waves are propagating from right to left. From the figure (b), orientation relationships of α↔ε transition are $[111]_{BCC}||[\bar{1}2\bar{1}0]_{HCP}$ and $[10\bar{1}]_{BCC}||[0001]_{HCP}$. Besides, stacking sequence of $\{0001\}_{HCP}$ (and $\{111\}_{FCC}$) are marked in (b).

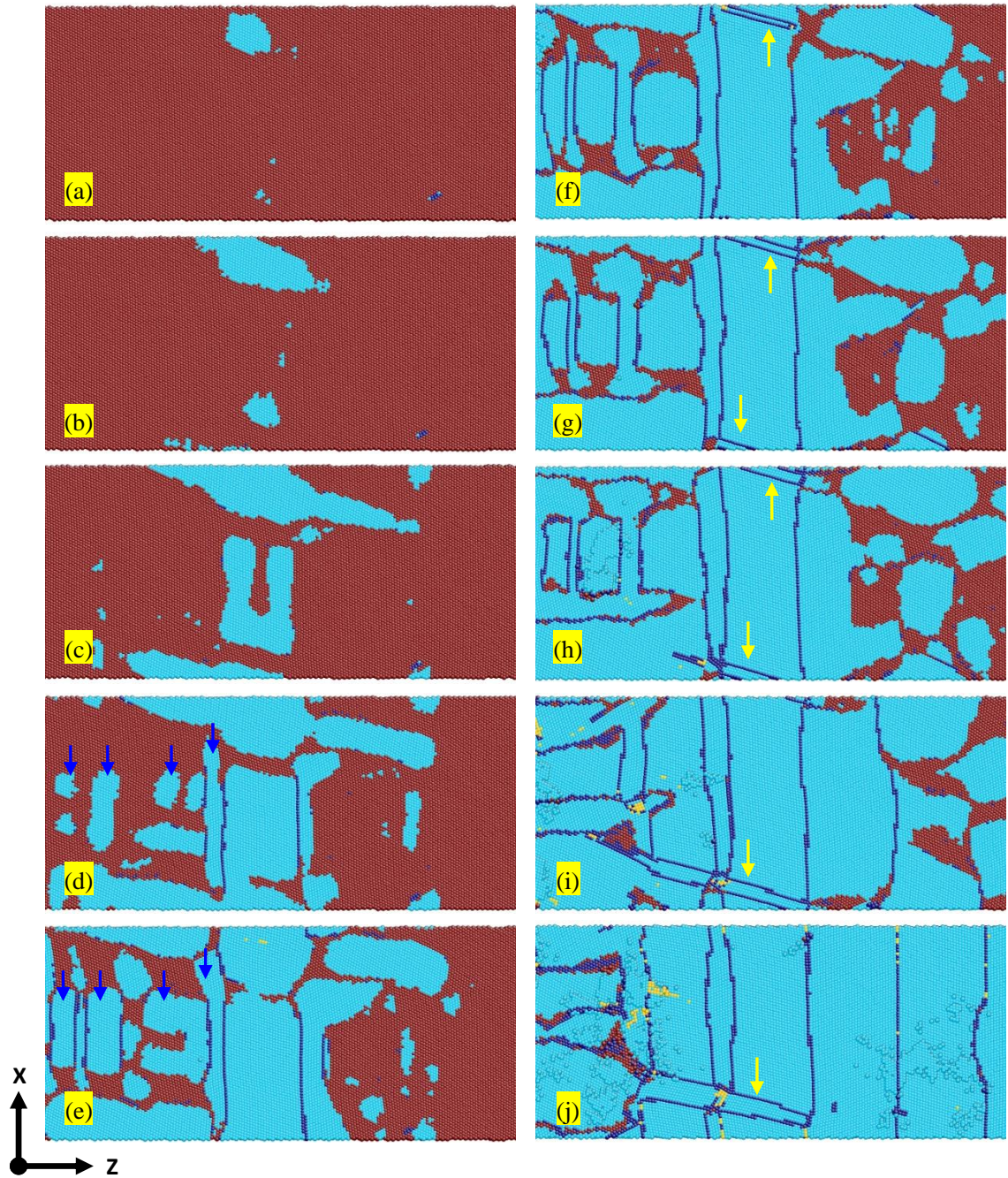

Fig. 4. Formation processes of reverse transition induced twins, as well as secondary twins, in sample I during recovery from the loading with $v_p^{max}$ = 0.4 km/s, where (a)-(j) corresponds to the moment of 65, 66, 67,68,69,70,71,72,75 and 80 ps, respectively. Atoms in the figures are colored by adaptive CNA method: red (HCP), yellow (FCC), light blue (BCC) and dark blue (defect atoms).

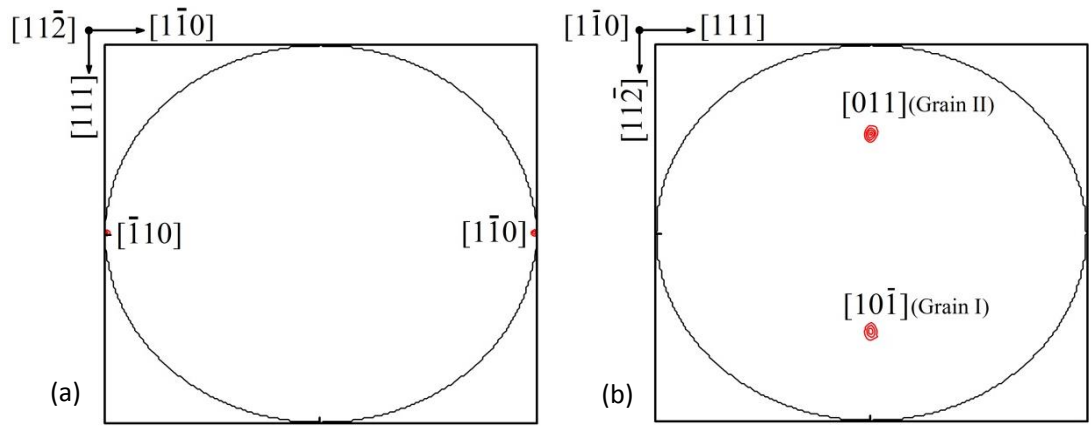

Fig. 5. (0001)$_{HCP}$ pole figures of (a) sample I after ramp compressions with $v_p^{max}$ = 0.8 km/s for 40ps, and (b) sample II under ramp compressions with $v_p^{max}$ = 0.8 km/s for. 30ps, where the sample reference frames are represented by the crystallographic reference frames of Grain I.

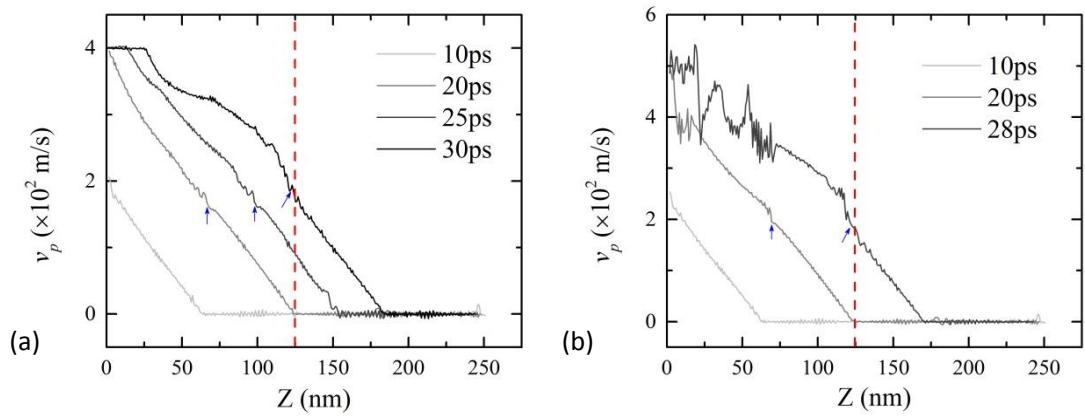

Fig. 6. Particle velocity profiles in sample I under loadings with (a) $v_p^{max}$ = 0.4 km/s and (b) $v_p^{max}$ = 0.5 km/s, where position of CTB is marked by the red dashed line. The blue arrows have marked out the position of the jump in elastic precursor.

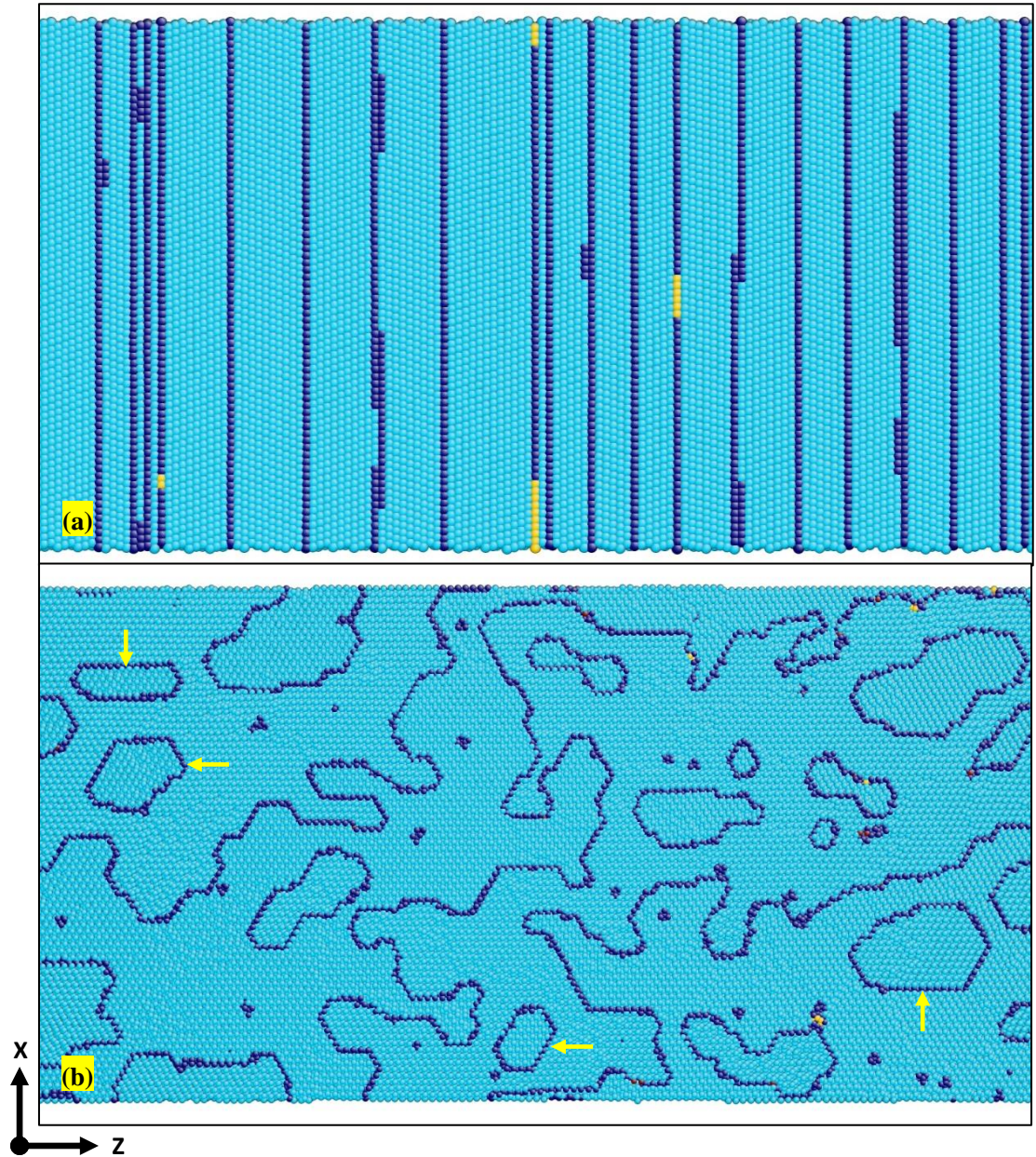

Fig. 7. (a) Sample I and (b) sample II recovered from the loading with $v_p^{max}$ = 0.4 km/s and $v_p^{max}$ = 0.5 km/s, respectively. Typical hexagon-shaped twins form after the reverse transition. Color scheme for atoms in the figures is the same as that in Fig. 4.

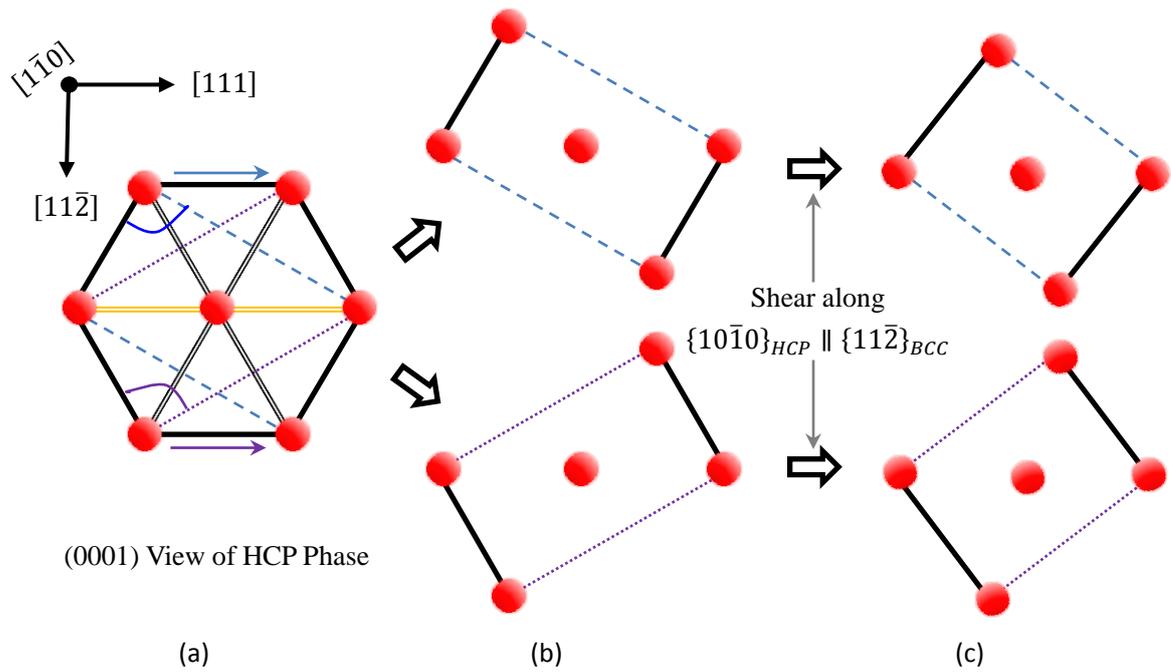

Fig. 8. Schematics showing the formation of twins via reverse transition from HCP to BCC phase under the unloading. The crystallographic reference frame shown in the figure corresponds to the one in Grain I of initial iron sample.

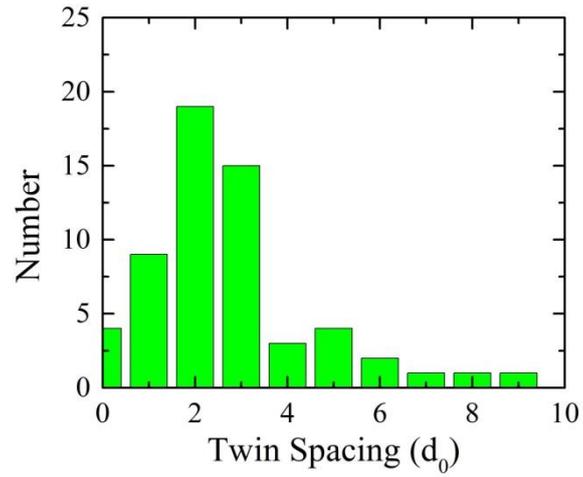

Fig. 9. Distribution of twin spacing in sample I after unloading for the case of $v_p^{max}$ = 0.4 km/s, where the twin spacing has been reduced by the separation between adjacent {112} planes (~ 7.0 Å).

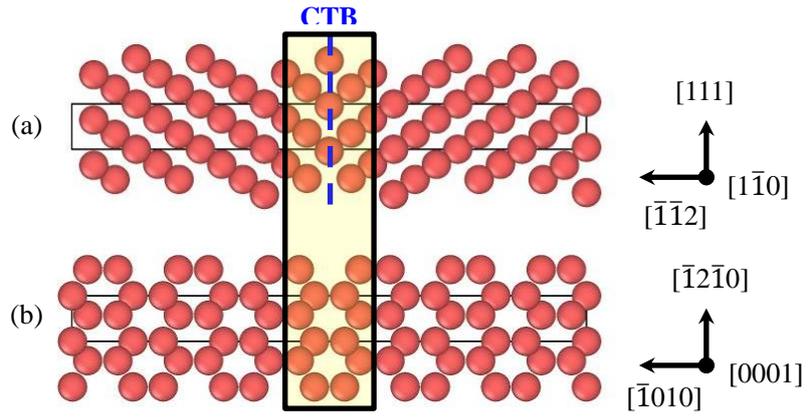

Fig. 10. (a) Initial and (b) finial configurations of transformation from BCC twin to HCP phase, where the light yellow plane marks the region of CTBs. Twin spacing is $2d_{\langle 112\rangle}$. Reference frame in (a) is represented by crystallographic orientations in the grain at right hand side of the CTB. The black rectangular corresponds to the minimum cell that keeps the twin spacing.

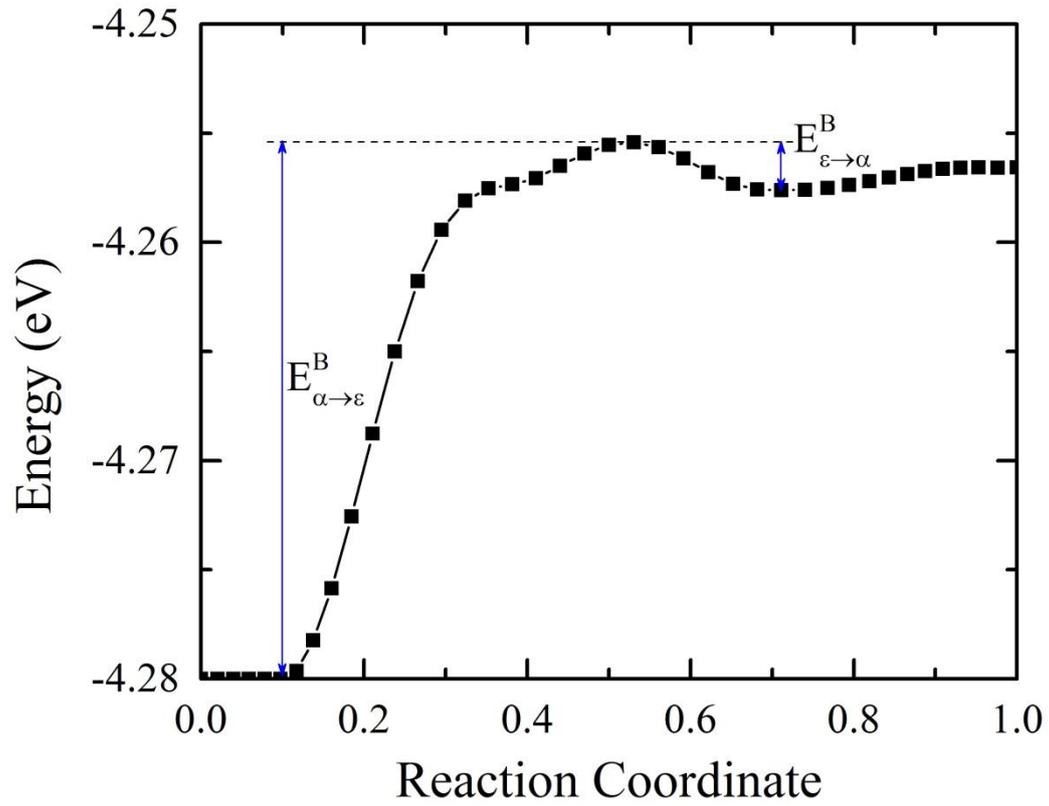

Fig. 11. Potential energies along minimum energy path (MEP) between BCC and HCP phase of single crystalline iron, where $E^B_{\alpha\to\varepsilon}$ and $E^B_{\varepsilon\to\alpha}$ represent energy barriers of the direct and reverse phase transition of iron, respectively.

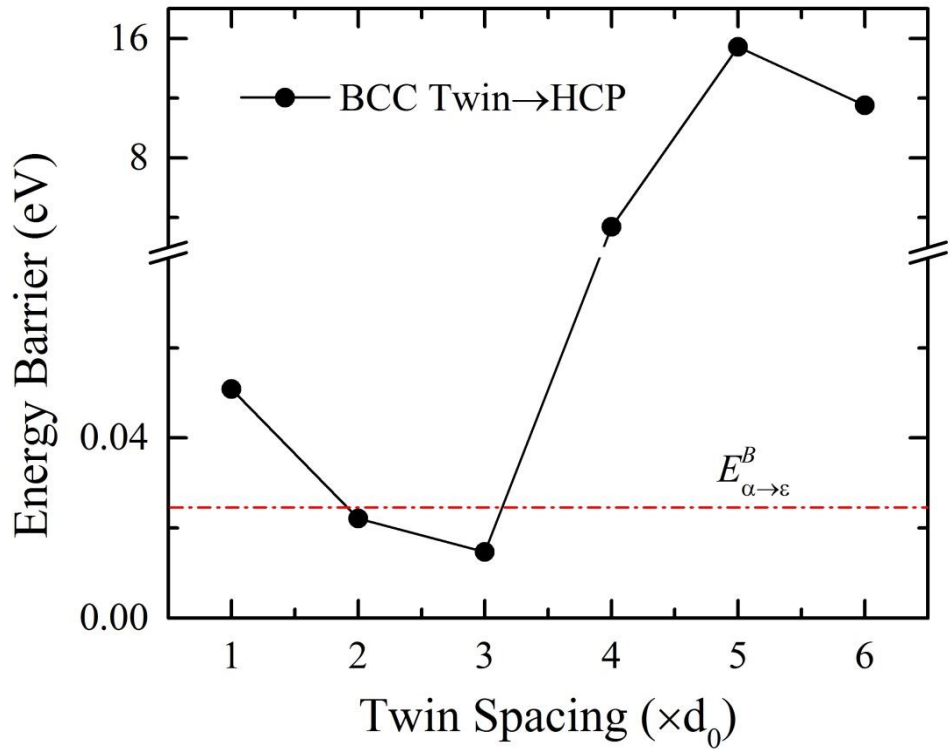

Fig. 12. Calculated energy barrier of transition from BCC twin to HCP phase as a function of twin spacing. The horizontal red dashed line corresponds to the energy barrier of the transition from BCC to HCP phase.

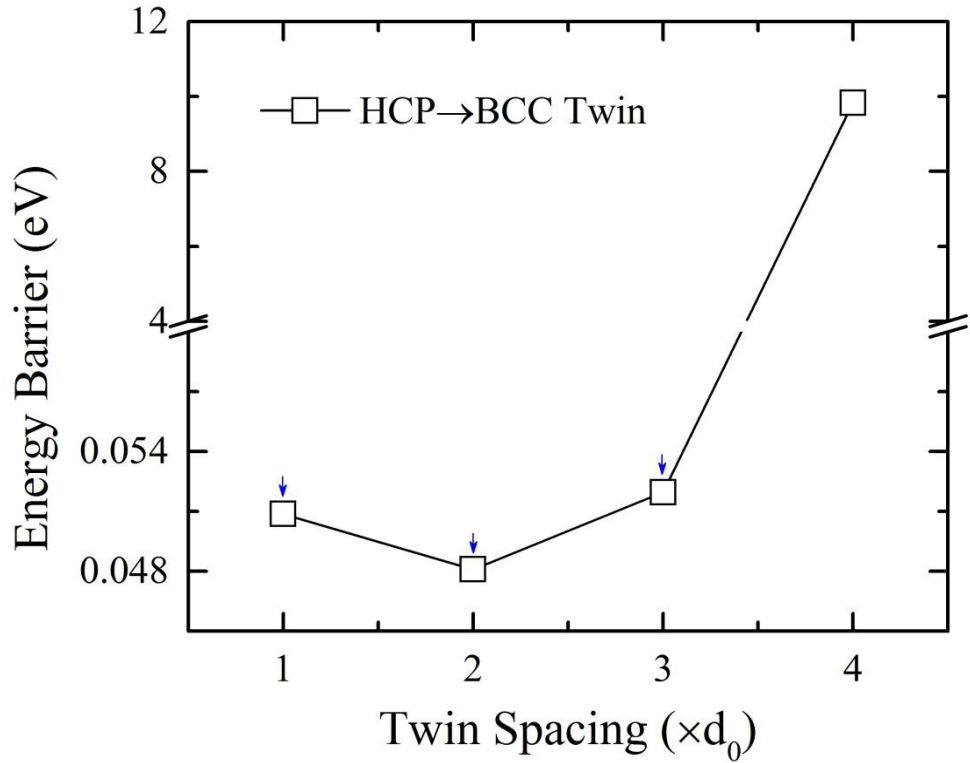

Fig. 13. Calculated energy barriers of transition from HCP phase to BCC twin in $E^d(R)$ curves defined by Eq. (8) for several twin spacing. The blue arrows mark the major twins observed after the reverse phase transition.